\documentstyle[psfig,aps]{revtex}
\tightenlines
\begin{document}
\draft
\preprint{\vbox{\it Submitted to Phys. Rev. D \hfill\rm CU-NPL-1153}}

\begin{title}
{Renormalization Group Flow Equations for the Sigma Model\cite{work}}
\end{title}
\author{A. S. Johnson\cite{lockheed} and J. A. McNeil}
\address{
Department of Physics, Colorado School of Mines, Golden, CO 80401}
\author{J. R. Shepard}
\address{
Department of Physics, University of Colorado, Boulder, CO 80309}
\date{\today}
\maketitle

\begin{abstract}
We present a nonperturbative renormalization group solution of the 
Gell-Mann--Levy $\sigma$-model
 which was originally proposed as a phenomenological description of the 
dynamics of nucleons and mesons.  In our version of the model the fermions 
are interpreted as quarks which interact via the $\sigma$ and $\pi$ mesons.  
We derive and numerically solve renormalization group (RG) flow equations 
to leading order in a derivative expansion to study the behavior of the model 
as it evolves from high to low momentum scales. We develop an expansion in 
chiral-symmetry-breaking which enables us to track this symmetry breaking
with the evolution of the scale. We use infrared observables to 
constrain the phenomenology allowing predictions of other quantities such as 
$\pi-\pi$ scattering lengths.  The results show improvement over the tree 
level calculation and are consistent with experiment and the results of 
alternate theoretical approaches such as chiral perturbation theory and 
lattice gauge theory. 
\end{abstract}

\vspace{1cm}

\parskip=2mm

\section{Introduction}
\label{intro}

The complete elucidation of nuclear dynamics in terms of 
Quantum Chromodynamics (QCD) is still forthcoming.  This elucidation 
involves a meeting in the middle of two complementary efforts.  On the one 
hand, using mainly symmetry patterns and the wealth of low energy 
experimental data, nuclear physicists have been able to construct models
valid for different regimes and deduce relationships between these models.  
On the other hand, since the early 1970s it has become increasingly clear 
that nuclear processes are the emergent phenomena of the dynamics of 
quarks and gluons (QCD).  That these threads can be pursued independently 
has been demonstrated by the quantitative success of nuclear models without 
any knowledge of QCD and the subsequent widespread 
acceptance of QCD largely on the basis 
of high energy behavior that appears to have little or no bearing on 
nuclear phenomena.  
But in the middle there ought to be some confluence.

Indeed, though the details are still murky, QCD does dictate that quarks 
and gluons are {\it confined} into hadrons within a region of about 1 fermi.  
Thus the QCD degrees of freedom at lower energies are transformed into 
the familiar nuclear degrees of freedom.  One can then begin with a model of 
quarks and mesons at the confinement scale and ask what relationship each 
of the parameters has with QCD parameters.  The form of the low energy 
dynamics is strongly constrained by {\it chiral symmetry} and its breaking 
in the hadronic sector.  The small departure from chiral symmetry is a 
consequence of the smallness of the up and down quark masses  with 
respect to the QCD scale $\Lambda_{QCD}\approx 200$-$300MeV$.  This can be 
exploited to construct an effective field theory for low energy QCD or 
what has been termed {\it chiral perturbation theory} ($\chi$PT).  
Here the picture is based on a sigma model whose parameters are 
determined phenomenologically.  The question arises, however, how to 
relate the parameters of the effective low energy theory to those of 
short distance QCD.  Intermediate to addressing this problem it is a 
desideratum to compute the scale dependence of parameters of simple 
models which incorporate the chiral symmetry breaking patterns of QCD.

In the present work we solve the Gell-Mann--Levy $\sigma$-model \cite{gml} 
nonperturbatively using Renormalization Group (RG) flow equations
where projections onto {\it momentum independent} 
couplings at each momentum step are made \cite{hh}.  This is equivalent to 
truncating the derivative expansion to  
leading order (LO), an approximation sometimes referred to as the ``local 
potential approximation''.  
A small quark mass is introduced at the ultraviolet scale to 
break chiral symmetry. The free parameters of the model can be constrained by 
infrared observables allowing modest predictive power.  The main result
 of the calculations, detailed below, are predictions of $\pi-\pi$ 
scattering lengths which 
are significant improvements over tree-level calculations and are 
consistent with experiment as well as with lattice and chiral 
perturbation theory results \cite{dinko}.  The present paper 
summarizes many of the results reported in Ref. \cite{mythes}.

The organization of this paper is as follows.  In section \ref{rev} we briefly 
outline the history of the RG approach to field theories.  In section 
\ref{der}, our derivation of the RG flow equations for the $\sigma$-model
is sketched.  The numerical technique for solving the flow equations is 
then outlined, our approach to constraining the parameters are discussed, 
and the results presented in section \ref{num}.  A summary and 
some extensions to the model are considered in section \ref{con}.  Three 
technical appendices include derivations of results used in the text.   

\section{Review of the RG Flow Equation}
\label{rev}

The renormalization group equation upon which this work is based was first derived by 
Wilson and Kogut \cite{wk} 
and Wegner and Houghton \cite{wh}
in two different but equivalent ways.  Wilson and Kogut derived their equation 
using a so-called ``smooth cutoff'' between integrated and unintegrated 
modes.  Their rationale was to avoid nonlocalities that naturally arise in 
position space when one makes sharp divisions in momentum space.  The price 
paid for this approach is that unconstrained smoothing functions must be introduced to 
facilitate the integrations.  These smoothing functions complicate the 
equations and make their solution---even after drastic approximations 
({\it e.g.} the leading order (LO) in the derivative expansion.)
---impossible analytically. There are also unphysical dependencies on the detailed
form of the smoothing function.  The so-called ``sharp-cutoff'' method of 
Wegner and Houghton \cite{wh} is simpler in that there are no smoothing 
functions to contend with, but nonlocal effects are now present. 
Another problem with this approach is that there appear to be 
ambiguities in the form of the RG equations that depend on a variable change 
at a particular point in the derivation \cite{morrisc}.
However, at least to leading order (LO) in the derivative expansion,
there is no ambiguity with the sharp cutoff approach which we 
employ in this paper.

One can imagine many different ways to approximate the exact RG equations.  A 
natural approximation scheme involves expanding the action in powers of 
momentum, or---in real space---in powers of derivatives of the fields.  After 
integrating over field components corresponding to momenta in the UV momentum shell, the 
leading order (LO) approximation is then obtained by setting all of the 
remaining components equal to zero except for the uniform component.
  (This is sometimes called the ``local potential 
approximation'' (LPA).)  The next to leading order (NLO) result is obtained 
by keeping one small momentum and so on.  The LO approximation to the full RG 
equations appears first in the work of Nicoll {\it et al.} \cite{nichst}.  

Hasenfratz and Hasenfratz \cite{hh}, in an early seminal work, showed that 
the RG equations approximated using a method equivalent to LO in the derivative 
expansion give interesting and nontrivial results.  
They solve for the RG flows of the effective potential in a pure scalar 
theory and extract critical exponents to compare with calculations 
performed by other means.  For $d=3$ they find impressive 
agreement for the critical exponents $\nu$ and $\omega$.  
Also for $d=4$ they find no nontrivial fixed point solution which is 
consistent with 
the triviality of $\phi^4$ field theories.  
Their work showed that even the  LO approximation gives 
a rich quantitative description of the critical properties of strongly-coupled scalar 
field theories. 

There has been a recent resurgence of interest in the RG effective action 
approach to scalar field theories.  Hasenfratz and Nager \cite{hn} study the 
cutoff dependence of the Higgs mass using RG methods.  Wetterich and 
co-workers \cite{wetta,tetwett,jungwett}, essentially extend the work of 
Hasenfratz and Hasenfratz to NLO use a smooth cutoff procedure.  
This allows them to include the effects of wavefunction 
renormalization and compute the critical exponent $\eta$ which they find to be
 in rough agreement with other calculations.  
Morris \cite{morrisa} studies approximations to the exact RG as derived using 
both smooth and sharp cutoffs and in Ref.\cite{morrisb} computes critical exponents 
to NLO noting that the scheme appears to converge.  In 
Ref.\cite{morrisc} the so-called sharp cutoff ambiguities are treated in 
detail and the DE of the effective action is reviewed in \cite{morrisd}.  
Alford \cite{alfa}, with an eye toward the electroweak phase transition, 
computes exponents with a sharp cutoff procedure and discusses some of the 
practical difficulties involved with extending the calculations to NLO.  
In an extensive RG and Monte Carlo analysis of general scalar field theories, 
Shepard {\it et al.} \cite{shepa} derive ``latticized'' RG flow equations in 
LO and demonstrate impressive aggreement with Monte Carlo results for a wide variety 
of cases, broken and unbroken phases, three and four dimensions, and O(1) and O(2) 
theories.   Some recent related work appears here as  Refs. \cite{liag}, \cite{ball}, \cite{halhu}.

In addition the inclusion of fermions has been addressed 
in the literature recently.  Maggiore \cite{magg} includes fermions 
via a generalized Yukawa term and 
derives LO flows for the scalar and generalized Yukawa potentials.  He finds 
no evidence that the fixed point structure of the theory is affected by the 
addition of fermions.  In a more comprehensive study, Clark and 
co-workers \cite{clarka} derive the exact RG equations for theories of 
arbitrary field content and derive from this equation the LO flows for 
the scalar and generalized Yukawa potentials which agree with the results of 
Ref.\cite{mythes,magg}.  In two related papers Clark {\it et al.} 
study the issue of 
computing mass bounds for scalars and fermions in the standard model 
\cite{clarkb} and the stability of fine tuned hierarchies \cite{clarkc}.  
In related work, Ellwanger and Vergara \cite{ellver} use RG flow equations 
for generalized NJL models to study the Higgs top quark system to leading 
order in $1/N_c$.  Other work pertaining to the Higgs top system appears 
as Ref.\cite{ahasena}.  

In an effort paralleling this work Jungnickel and Wetterich \cite{jungwett} have 
applied renormalization group methods to the generalized $\sigma$-model. In one 
sense their work is more 
ambitious than the present work.  They have considered the general SU(N) sector
with all allowable functions of the field invariants and included via a smooth 
cut-off the NLO quantities which allow treatment of the anomalous 
dimension.  But the price
paid for such generality is an abundance of free parameters, some non-physical,
 which are difficult to 
constrain.  In contrast, the present work focuses exclusively on the SU(2) 
sector which requires only one field invariant, uses a sharp cut-off approach 
as in Ref.\cite{hh} truncated at leading order in the derivative expansion. 
While this precludes any treatment of the strange-quark sector or the anomalous 
dimension, there are fewer parameters which can be well constrained by 
experimental data allowing 
 modest predictive power for the theory.

There is now ample evidence in the literature that the leading order approxiamtion 
to the RG flow of a field theory incorporates correctly many of the 
intricacies of strong-coupled quantum field theories.   The aim of the 
present work is to apply this powerful method to the 
the Gell-Mann--Levy $\sigma$-model in an attempt to understand better low energy 
nuclear phenomenology.
      
\section{Derivation of the Flow Equations}
\label{der}
\subsection{General}

We first present a derivation of an exact RG equation for an arbitrarily 
complicated quantum field theory described by the action, 
$S^{(\Lambda)}[\Phi]$, regulated at some large momentum cutoff $\Lambda$.
  The field content of the theory can in principle be anything at all, 
$\Phi=\{\phi, \psi, A_\mu, F_{\mu \nu}, \dots\}$ {\it i.e.} fields 
described by complex scalars, spinors, vectors and/or tensors, 
though we will only be 
concerned with theories containing scalars and spinors.  
We define the effective action at the 
momentum scale $\Lambda-\Delta\Lambda$, $S^{(\Lambda-\Delta\Lambda)}$ as,
\begin{equation}
     {\rm e}^{-S^{(\Lambda-\Delta\Lambda)}}=
     \int_{shell} {\cal D}\overline\Phi{\cal D}\Phi\ 
     {\rm e}^{-S^{(\Lambda)}},
     \label{finta}
\end{equation}
where the subscript ``{\it shell}'' indicates that only the Fourier 
components of the fields \{$\Phi_q$\} with momenta in the shell, 
$\Lambda-\Delta\Lambda<
|q|<\Lambda$ are integrated.  Thus the notation for the measure means,
\begin{equation}
     {\cal D}\overline\Phi{\cal D}\Phi=\prod_{q_1}d\overline\Phi_{q_1}
     \prod_{q_2}d\Phi_{q_2},
\end{equation}
for $\Lambda-\Delta\Lambda<|q_1|,|q_2|<\Lambda$.  $\overline\Phi$ is the 
``conjugate'' of $\Phi$ ({\it e.g.} complex conjugate for the complex scalar 
field $\phi$, Dirac adjoint for the Dirac spinor $\psi$, Hermitian conjugate 
for matrix fields {\it etc.}).  The 
actions $S^{(\Lambda-\Delta\Lambda)}$ and $S^{(\Lambda)}$ are considered 
{\it equivalent} in the sense that for $|q|\ll\Lambda$ they each give the 
same $n$-point functions.  We now decompose $\Phi(x)$ into uniform and 
nonuniform pieces,
\begin{eqnarray}
     \Phi(x)&=&\Omega_0+\Omega(x) \nonumber \\
     \Omega(x)&=&\sum_{q\not=0}\Omega_q{\rm e}^{iq\cdot x}.
     \label{fdec}
\end{eqnarray}
Expanding $S^{(\Lambda)}$ in a functional Taylor series about the uniform 
field components gives,
\begin{eqnarray}
     S^{(\Lambda)}[\overline\Phi,\Phi]&=&
     S^{(\Lambda)}[\overline\Omega_0,\Omega_0]\biggl\vert_0
     +{\sum_{q\not=0}}'\biggl[{\delta S\over{\delta\Omega_q}}\biggl\vert_0
     \Omega_q+
     \overline\Omega_q{\delta S\over{\delta\overline\Omega_q}}\biggl\vert_0
     \biggr] 
     \nonumber \\
     && \qquad \qquad \qquad +{\sum_{q_1,q_2\not=0}}'\ \overline\Omega{q_1}
     {\delta^2 S \over{\delta\overline\Omega_{q_1}\delta\Omega_{q_2}}}
     \biggl\vert_0\Omega{q_2} \nonumber \\
     &=& S^{(\Lambda)}\biggl\vert_0+\overline{\cal J}\Omega+
     \overline\Omega{\cal J}+\overline\Omega M\Omega.
     \label{acta}
\end{eqnarray} 
The subscript zero means that all modes, $\Omega_q$, with $q$ in the 
shell are set to zero; and the primed summation symbol 
indicates that only momenta in the 
shell are included in the sum.  For the purposes of deriving differential 
equations with respect to the independent variable $\Lambda$, we will 
eventually take the limit $\Delta\Lambda\rightarrow 0$, thus we can 
truncate the series after the quadratic term {\it without approximation} since 
all higher order contributions to the RG flows will be  at least 
${\cal O}(\Delta\Lambda^2)$.  
In the second equality several definitions have been made:
\begin{eqnarray}
     \overline{\cal J}\equiv\pmatrix{
     {\delta S\over{\delta\Omega_{q_1}}}\bigl\vert_0\cr
     {\delta S\over{\delta\Omega_{q_2}}}\bigl\vert_0\cr
     \vdots},
\end{eqnarray}
and its conjugate are the ``generalized source'' column vectors and the 
supermatrix of second derivatives is defined as (see  Appendix A
 for a review of the supermatrix formalism),
\begin{eqnarray}
     M\equiv\pmatrix{\Sigma&{\cal A}\cr
     \overline{\cal A}&{\cal F}}.
\end{eqnarray}
Also, where momentum subscripts are not present, matrix multiplication 
over the momentum indices is implied, {\it e.g.},
\begin{eqnarray}
     \overline{\cal J}\Omega&=&\pmatrix{
     {\delta S\over{\delta\Omega_{q_1}}}\bigl\vert_0&
     {\delta S\over{\delta\Omega_{q_2}}}\bigl\vert_0 & \ldots}
     \pmatrix{\Omega_{q_1}\cr
     \Omega_{q_2}\cr
     \vdots} \nonumber \\
     &=&\sum_{q\not=0}{\delta S\over{\delta\Omega_{q}}}\biggl\vert_0
     \Omega_{q}.
\end{eqnarray}

Now substituting Eq.(\ref{acta}) into Eq.(\ref{finta}) gives,
\begin{eqnarray}
     {\rm e}^{-S^{(\Lambda-\Delta\Lambda)}}&=& 
     {\rm e}^{-S^{(\Lambda)}\vert_0}\int {\cal D}\overline\Omega
     {\cal D}\Omega\ {\rm e}^{-(\overline\Omega M\Omega+
     \overline{\cal J}\Omega+\overline\Omega{\cal J})}
     \nonumber \\
     &=&{\rm e}^{-S^{(\Lambda)}\vert_0} \{{\rm e}^{\overline{\cal J}
     M^{-1}{\cal J}} {\rm sdet}^{-1} M\}.
\end{eqnarray}  
Irrelevant constant factors have been dropped wherever they appear.  Using 
the identity,
\begin{equation}
     {\rm sdet}^{-1}M={\rm e}^{-{\rm str}\ {\rm ln}\ M},
\end{equation}
the exact RG equation for this generalized system is easily obtained as,
\begin{equation}
     S^{(\Lambda-\Delta\Lambda)}=S^{(\Lambda)}\biggl\vert_0
     +\ {\rm str}\ {\rm ln}\ M-\overline{\cal J}M^{-1}{\cal J}
     +{\cal O}(\Delta\Lambda^2).
     \label{erga}
\end{equation}   
This equation relates the action at momentum scale $\Lambda-\Delta\Lambda$ to 
the action at momentum scale $\Lambda$.  We will not convert it directly into 
a  functional differential equation (as Clark {\it et al.} \cite{clarka} do) 
until we 
consider its form for a particular action.  A 
terminology has grown up around equations such as (\ref{erga}).  The 
``${\rm str\ ln}$'' term is generally referred to as the ``loop'' term and the 
``$\overline{\cal J}M^{-1}{\cal J}$'' term is referred to as the ``tree'' 
term.  This is because the contribution from the 
$\overline{\cal J}M^{-1}{\cal J}$ term is present in mean field theory 
whereas the contribution from the ${\rm str\ ln}$ term includes effects 
from loop integrations \cite{polch}.

For the purposes of treating the $\sigma$-model we consider only real 
scalar and spinor fields ({\it i.e.} $\Phi=\{\phi,\psi\}$). If the flow equations are 
truncated to LO in the derivative expansion, many simplifications ensue.  The 
source column vector becomes,
\begin{eqnarray}
     \overline{\cal J}=\pmatrix{\overline J \cr \overline\eta}
     =\pmatrix{J^*_{q_1}\cr J^*_{q_2}\cr \vdots \cr - \cr  
     \overline\eta_{q_1}\cr \overline\eta_{q_2}\cr \vdots}
     \label{soloa}
\end{eqnarray}
where,
\begin{eqnarray}
     J^*_q&\equiv&{\delta S\over{\delta\phi_q}}\biggl\vert_0 \nonumber \\
     \overline\eta_q&\equiv&{\delta S\over{\delta\psi_q}}\biggl\vert_0.
\end{eqnarray}
The matrix of second derivatives is,
\begin{eqnarray}
     M=\pmatrix{\Sigma&{\cal A}\cr
     \overline{\cal A}&{\cal F}}
     =\pmatrix{{\delta^2 S \over{\delta\phi_{-q}
     \delta\phi_{q}}}
     \biggl\vert_0&
     {\delta^2 S \over{\delta\overline\psi_{-q}
     \delta\phi_{q}}}\biggl\vert_0 \cr
     {\delta^2 S \over{\delta\phi_{-q}
     \delta\psi_{q}}}
     \biggl\vert_0& 
     {\delta^2 S \over{\delta\overline\psi_{-q}
     \delta\psi_{q}}}
     \biggl\vert_0}.
     \label{mloa}
\end{eqnarray}
Now we can rewrite the superdeterminant appearing in Eq.(\ref{erga}) 
using some 
tricks from Appendix \ref{appendixa} (compare with Eq.(\ref{appa})),
\begin{equation}
     {\rm sdet}^{-1}M=({\rm det}^{-1/2}N)({\rm det}{\cal F})
\end{equation}
(where $N\equiv\Sigma-\overline{\cal A}{\cal F}^{-1}{\cal A}$) which gives,
\begin{equation}
     {\rm strln}M={1\over 2}{\rm trln}N-{\rm trln}{\cal F}.
\end{equation}
(The $1/2$ coming from the counting in the momentum sums due to the 
{\it real} $\phi$.)  Now Eq.(\ref{erga}) becomes,
\begin{equation}
     S^{(\Lambda-\Delta\Lambda)}=S^{(\Lambda)}\biggl\vert_0
     +\ {1\over 2}{\rm trln}N-{\rm trln}{\cal F}
     -\overline{\cal J}M^{-1}{\cal J}.
     \label{ergaa}
\end{equation}

\begin{figure}
\centerline{\psfig{figure=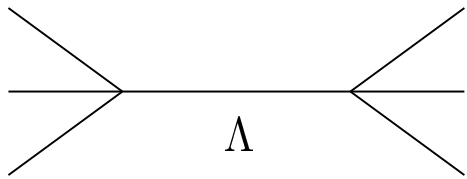,height=1in,width=2.5in}}
\caption{Tree diagram}
\label{tree}
\end{figure}

Restricted to LO in derivatives the last
(``tree'') term in this equation may be neglected.  One way to see this
 diagrammatically is to note that any particular contribution to the tree 
term in Eq.(\ref{erga}) or (\ref{ergaa}) will be comprised of external
 legs connected by a single {\it internal} propagator with momentum in 
the shell (see Fig. \ref{tree}).  Since to LO we set the 
momentum of all external legs to zero, these diagrams cannot conserve 
momentum at their vertices and therefore vanish.  
To make the argument analytically we write a particular term of $S'_q$,
\begin{equation}
{\delta S\over{\delta\phi_{q'}}}\biggl\vert_0=\cdots+{1\over{2!}}V'''
     \sum_{q_1,q_2\not=0}\phi_{q_1}\phi_{q_2}\delta_{q_1+q_2+q',0}+\cdots,
\end{equation}
where $q'$ is in the shell.  When, in going to LO we set all field components, 
$\phi_q$, with nonzero momentum to zero, it's clear that this term and all 
higher order terms will vanish.  Thus we write Eq.(\ref{ergaa}) as,
\begin{equation}
     S^{(\Lambda-\Delta\Lambda)}=
      S^{(\Lambda)}\biggl\vert_0
     -{1\over{2}}{\rm trln}N+{\rm trln}{\cal F}.
     \label{rga}
\end{equation}
This is the form of the RG flow equation that we use for our 
calculation in the present work.

\subsection{$\sigma$-model Flow Equations}
\label{dersig}

We now apply the results of the previous section to derive the flow equations 
for the $\sigma$-model.  The incorporation 
of chiral symmetry breaking brings a substantial price in the complexity 
of the algebra.  Many of the tedious details are relegated to the Appendices.  
Ref.\cite{mythes} includes many details not related in this paper.   

The Euclidean action for the $\sigma$-model is:
\begin{equation}
     S^{(\Lambda)}=\int d^dx\ \biggl[V^{(\Lambda)}(\rho,\sigma)+
     {1\over2}(\partial_\mu\phi^a)^2+
     \overline\psi[{\partial\llap{\slash}}+
     U^{(\Lambda)}(\rho,\sigma,{\cal G})]\psi\biggr].
     \label{sigact}
\end{equation}
A few definitions to compactify the notation have been made:
\begin{eqnarray}
     \phi^a&=&\pmatrix{{\sigma}\cr{\pi^i}}, \qquad a=0,i; \qquad 
               i=1,2,\dots, N \nonumber \\
     \Gamma^a&=&\pmatrix{{1}\cr{i\gamma^5\tau^i}}, \nonumber \\
     {\cal G}&=&\Gamma^a\phi^a=\sigma+i\gamma^5\vec\tau\cdot\vec\pi,
          \nonumber \\
     \rho^2&=&{\cal G}{\cal G}^{\dagger}=(\phi^a)^2=\sigma^2+(\vec\pi)^2,
\end{eqnarray}
where the last term is the only field invariant when specifying $\Gamma^a$ 
only to the $N=4$ or $O(4)$ case. 
 We will often use the more general notation but will only be considering 
$O(4)$ in this work.  Recall that $SU(2)\times SU(2)$ rotations of the 
fermions induce $O(4)$ rotations in the boson sector.  The scalar sector 
of the $\sigma$-model then is a four component field.  The potentials 
$V^{(\Lambda)}(\rho,\sigma)$ and $U^{(\Lambda)}(\rho,\sigma,{\cal G})$ 
include terms proportional to powers of the zeroth component of the scalar 
field, $\sigma$, which will break chiral symmetry.  To keep track of the 
symmetry breaking as the theory flows from high to low momentum we expand 
the potentials in the form,
\begin{eqnarray}
     V(\rho,\sigma)&=&V_0(\rho)+\sigma V_1(\rho)
     +{\sigma^2\over{2}}V_2(\rho)+\cdots 
     \nonumber \\
     U(\rho,\sigma,{\cal G})&=&m(\rho,\sigma)+{\cal G} g(\rho,\sigma)
     \label{sigpots}
\end{eqnarray}
where $m(\rho,\sigma)$ and $g(\rho,\sigma)$ are expanded similarly to 
$V(\rho,\sigma)$.  Each of the functions $V_k$, $m_k$ and $g_k$ for $k=0,1,2$ 
is $SU(2)\times SU(2)$ symmetric since they only depend on $\rho$.  Chiral 
symmetry is broken by $V_1$, $V_2$, $g_1$, $g_2$ and $m(\rho,\sigma)$.

The LO flow of the action Eq.(\ref{sigact}) is given by Eq.(\ref{rga}) 
rewritten as,
\begin{equation}
     S^{(\Lambda-\Delta\Lambda)}=
      S^{(\Lambda)}\biggl\vert_0
     -{\rm trln}{\bf N}+{\rm trln}{\bf\cal F}.
     \label{rgamod}
\end{equation}
where ${\bf N}$ has components $N^{ab}$,
\begin{eqnarray}
     N^{ab}&\equiv&\Sigma^{ab}-\overline{\cal A}^a {\cal F}^{-1} {\cal A}^b
          \nonumber \\
     \\ \nonumber
     \pmatrix{\Sigma&{\cal A}\cr
     \overline{\cal A}&{\cal F}}
     &=&\pmatrix{{\delta^2 S \over{\delta\phi^a_{-q}
     \delta\phi^b_{q}}}
     \biggl\vert_0&
     {\delta^2 S \over{\delta\overline\psi_{-q}
     \delta\phi^b_{q}}}\biggl\vert_0 \cr
     {\delta^2 S \over{\delta\phi^a_{-q}
     \delta\psi_{q}}}
     \biggl\vert_0& 
     {\delta^2 S \over{\delta\overline\psi_{-q}
     \delta\psi_{q}}}
     \biggl\vert_0}.
     \label{mats}
\end{eqnarray}
The extra indices now refer to the components of the scalar field.  We 
Taylor expand the potentials,
\begin{equation}
     V^{(\Lambda)}(\phi^a)=V^{(\Lambda)}(\phi_0^a)
     +V^{'(\Lambda)a}(\phi_0^a)\varphi^a
     +{1\over 2}V^{''(\Lambda)ab}(\phi_0^a)\varphi^a\varphi^b+\cdots
\end{equation}
(and similarly for $U$).  With normalization 
\begin{equation}
     \int d^dx\ {\rm e}^{iq\cdot x}=({\rm Vol})\delta_{q,0}
\end{equation} 
we get,
\begin{eqnarray}
     S^{(\Lambda)}\over{\rm Vol}&=&V^{(\Lambda)}(\phi_0^a)
     +\overline\psi_0U^{(\Lambda)}(\phi_0^a)\psi_0 
     \nonumber \\
     &&+\sum_{q\not=0}
     \biggl[{1\over2}\phi_{-q}^a\bigl[\delta^{ab}q^2
     +V^{''(\Lambda)ab}(\phi_0^a)
     +\overline\psi_0 U^{''(\Lambda)ab}(\phi_0^a)\psi_0\bigr]\phi_q^b
     \nonumber \\
     &&+\overline\psi_{-q}^\alpha\bigl[iq\llap{\slash}
     +U^{(\Lambda)}(\phi_0^a)\bigr]\psi_q^\alpha
     \nonumber \\
     &&+U^{'(\Lambda)a}(\phi_0^a)(\overline\psi_0\phi_{-q}^a\psi_q
     +\overline\psi_{-q}\phi_q^a\psi_0)\biggr]+\cdots\cdot
\end{eqnarray}
From this expression we can compute all the matrices in Eq.(\ref{rgamod}),
\begin{eqnarray}
     \Sigma^{ab}&=&\delta^{ab}q^2+V^{''ab}(\phi^a_0) \nonumber \\
     \Omega^{ab}&\equiv&U^{''ab}-{2U^{'a}U^{'b}\over{iq\llap{\slash}+U}}
          \nonumber \\
     {{\cal F}\over{{\rm Vol}}}&=&{1\over{2}}(iq\llap{\slash}+U) \nonumber \\
     {N^{ab}\over{\rm Vol}}&=&\Sigma^{ab}
               +\overline\psi_0\Omega^{ab}\psi_0.
\end{eqnarray}
With these definitions Eq.(\ref{rgamod}) becomes,
\begin{eqnarray}
     S^{(\Lambda-\Delta\Lambda)}\over{\rm Vol}&=&
     V^{(\Lambda-\Delta\Lambda)}(\phi_0^a)
     +\overline\psi_0U^{(\Lambda-\Delta\Lambda)}(\phi_0^a)\psi_0 
     \nonumber \\
     &=&V^{(\Lambda)}(\phi_0^a)
     +\overline\psi_0U^{(\Lambda)}(\phi_0^a)\psi_0 
     \nonumber \\
     &&-{1\over{\rm 2Vol}}{\rm tr}\ {\rm ln}\biggl[\Sigma^{ab}
     +\overline\psi_0\biggl(\Omega^{ab})\psi_0\biggr]
     \nonumber \\
     &&+{1\over{\rm Vol}}{\rm tr}\ {\rm ln}(i{q{\mkern -10.mu}{/}}
     +U^{(\Lambda)}(\phi_0^a))+\cdots\cdot
     \label{bigact}
\end{eqnarray}
We can write Eq.(\ref{bigact}) as two coupled equations for the flow of 
$V$ and $U$.  Before doing so, we make a number of modifications.  
First, we can write,
\begin{eqnarray}
     {1\over{iq\llap{\slash}+U}}&=&{iq\llap{\slash}+U\over{q^2+UU^{\dagger}}}\simeq
          {U^{\dagger}\over{D_F}}, \nonumber \\
     D_F &\equiv& q^2+UU^{\dagger}=q^2+m^2+2\sigma mg+\rho^2 g^2,
\end{eqnarray}
where we've dropped the term proportional to $q\llap{\slash}$ since it will vanish 
at LO when we take angle averages.  Second, we can write the trace over the 
inverse fermion propagator term as,
\begin{equation}
     {\rm tr}\ {\rm ln}(i q\llap{\slash}+U)=
          {1\over 2}n_fc_d\sum_q{\rm ln}D_F
\end{equation}
with $c_d=2^{d/2}(2^{(d-1)/2})$ for d even (odd), using the identity 
$|iq\llap{\slash}+U|^2=q^2+U^2$; the $n_fc_d\sum_q$ factor comes from the trace 
over the flavor and Dirac indices.  Also we can write,
\begin{equation}
     {\rm tr}\ {\rm ln}\biggl[\Sigma^{ab}+\overline\psi_0\biggl(\Omega^{ab}
     \biggr)\psi_0\biggr]
     ={\rm ln}\ {\rm det}\ \Sigma^{ab}+{\rm tr}\ [(\Sigma^{ac})^{-1}
     \overline\psi_0\Omega^{cb}\psi_0],
\end{equation}
where the trace in the last term is only over the $ab$ indices.  Putting all 
this together, we can write the $V$ and $U$ equations from Eq.(\ref{bigact}) as,
\begin{eqnarray}
     V^{(\Lambda-\Delta\Lambda)}&=&V^{(\Lambda)}
     -{1\over{2{\rm Vol}}}\sum_q({\rm ln}\ {\rm det}\ {\bf\Sigma}
     -n_fc_d\ {\rm ln}\ D_F)
     \nonumber \\
     U^{(\Lambda-\Delta\Lambda)}&=&U^{(\Lambda)}
     -{1\over{2{\rm Vol}}}\sum_q{\rm tr}\ {\bf\Sigma}^{-1}\cdot{\bf\Omega}.
\end{eqnarray}
Now taking the limit $\Delta\Lambda\rightarrow 0$ we have,
\begin{eqnarray}
     \Lambda{\partial V^{(\Lambda)}\over{\partial \Lambda}}&=&
     -{A_d\over{2}}\Lambda^d\ ({\rm ln}\ {\rm det}\ {\bf\Sigma}
     -n_fc_d\ {\rm ln}\ D_F)
     \label{flowa} \\
     \Lambda{\partial U^{(\Lambda)}\over{\partial \Lambda}}&=&
     -{A_d\over{2}}\Lambda^d\ {\rm tr}\ ({\bf\Sigma}^{-1}\cdot{\bf\Omega}),
     \label{chflows}
\end{eqnarray}
where the determinant and trace are only over the $ab$ indices.  In 
Eqs.(\ref{flowa}) and (\ref{chflows}),
\begin{eqnarray}
     \Sigma^{ab}&=&\delta^{ab}\Lambda^2+V^{''ab} \nonumber \\
     \Omega^{ab}&=&U^{''ab}-{2U^{'a}U^{\dagger}U^{'b}\over{D_F}}
          \nonumber \\
     D_F&=&\Lambda^2+m^2+\rho^2+2\sigma mg.
\end{eqnarray}
The matrices $V^{''ab}$, $U^{''ab}$, and $U^{'a}$ are worked out in terms of 
derivatives with respect to $\rho$ in Appendix B.  One glance 
at these expressions is adequate to impress the reader of the proliferation of 
algebraic complication in the extension of the simple Yukawa coupled fermions 
\cite{mythes,magg,clarka} to the case of broken chiral symmetry.  The matrices 
can be simplified by a similarity transformation, however, since the determinant 
and the trace are invariant with respect to such transformations.  As discussed 
in Appendix B, when similarity transformed, ${\bf \Sigma}$ has 
only six nonzero elements,
\begin{equation}
     {\bf \Sigma}^{'}={\bf S}\cdot{\bf \Sigma}\cdot{\bf S}^T
               =\pmatrix{\Sigma_{00} & \Sigma_{01} & 0 & 0 \cr
                    \Sigma_{10} & \Sigma_{11} & 0 & 0 \cr
                     0 & 0 & \Sigma_{22} & 0 \cr
                     0 & 0 & 0 & \Sigma_{33}},
     \label{bigsigpr}
\end{equation}
but ${\bf \Omega}^{'}$ still has every element nonzero although each of the 
elements is somewhat simpler.  Notice, however, that since only the 
combination ${\rm tr}{\bf \Sigma}^{'-1}\cdot{\bf \Omega}^{'}$ appears in 
the flow equations, only six elements of ${\bf \Omega}^{'}$ will contribute so that,
\begin{eqnarray}
     {\rm tr}\ {\bf \Sigma}^{'-1}\cdot{\bf \Omega}^{'}&=&
          (\Sigma^{00})^{-1}\Omega^{00}+
          (\Sigma^{10})^{-1}\Omega^{01}+
          (\Sigma^{01})^{-1}\Omega^{00}
     \nonumber \\
          &&+(\Sigma^{11})^{-1}\Omega^{11}+
          (\Sigma^{22})^{-1}\Omega^{22}+
          (\Sigma^{33})^{-1}\Omega^{33},
     \label{traceomsig}
\end{eqnarray}
where, {\it e.g.}, $(\Sigma^{00})^{-1}$ means the $00$ element of ${\bf \Sigma}^{'-1}$.

We can now consider the ``chiral limit'' where we set all chiral breaking 
terms ($m(\rho,\sigma)$, $V_1$, $V_2$, $g_1$ and $g_2$) to zero.  Then,
\begin{equation}
     \Sigma^{ab}\longrightarrow (\Lambda^2+V^{'}_0(\rho)/\rho)\delta^{ab}
          +(V^{''}_0(\rho)-V^{'}_0/\rho){\phi^a\phi^b\over{\rho^2}},
\end{equation}
which becomes, after a similarity transformation,
\begin{equation}
     \Sigma^{ab'}=\pmatrix{D_{\phi}&0&0&0 \cr
                    0&D_{\pi}&0&0 \cr
                    0&0&D_{\pi}&0 \cr
                    0&0&0&D_{\pi}},
     \label{bigsig}
\end{equation}
where $D_{\phi}=\Lambda^2+V^{''}_0$ and $D_{\pi}=\Lambda^2+V^{'}_0/\rho$.  Thus,
\begin{equation}
     (\Sigma^{ab'})^{-1}=\pmatrix{D_{\phi}^{-1}&0&0&0 \cr
                    0&D_{\pi}^{-1}&0&0 \cr
                    0&0&D_{\pi}^{-1}&0 \cr
                    0&0&0&D_{\pi}^{-1}},
     \label{bigsiginv}
\end{equation}
and so in the chiral limit only the diagonal elements of ${\bf \Omega}$ will 
contribute to the trace term.  These can be shown to be,
\begin{eqnarray}
     \Omega^{00}&=&{\cal G}[g^{''}_0+{2g^{'}\over{\rho}}-{2g_0\over{D_F}}
          (g^2_0+2g^2_0\rho^2+\rho^2g^{'2}_0)] \nonumber \\
     \Omega^{11}&=&\Omega^{22}=\Omega^{33}={\cal G}\bigl({g_0^{'}\over{\rho}}
          +{2g^3_0\over{D_F}}\bigr).
\end{eqnarray}
Thus from Eqs. (\ref{flowa}) and (\ref{chflows}) the $O(4)$ flow equations in 
the chiral limit are,
\begin{eqnarray}
     \Lambda{\partial V_0\over{\partial\Lambda}}&=&
          -{A_d\over{2}}\Lambda^d\bigl({\rm ln}\ D_{\phi}
          +3{\rm ln}\ D_{\pi}-n_fc_d\ {\rm ln}\ D_F\bigr)
     \label{vzero} \\
     \Lambda{\partial g_0\over{\partial\Lambda}}&=&
     -{A_d\over{2}}\Lambda^d\biggl\{{1\over{D_{\phi}}}\biggl[g^{''}_0
          +{2g^{'}\over{\rho}}-{2g_0\over{D_F}}
          (g^2_0+2g^2_0\rho^2+\rho^2g^{'2}_0)\biggr]
          \nonumber \\
          && \qquad \qquad+{3\over{D_{\pi}}}
          \biggl({g_0^{'}\over{\rho}}+{2g^3_0\over{D_F}}\biggr)\biggr\}.
     \label{gzero}
\end{eqnarray}
For $O(N)$ symmetric theories there would be an $N-1$ in place of the $3$s 
in front of the ${\rm ln}D_{\pi}$ and the $(1/D_{\pi})(\cdots)$ terms.  In 
Ref.\cite{mythes,hh,shepa} the flow equation for an $O(N)$ scalar-only field 
theory is derived and is equivalent to Eq.(\ref{vzero}) with $n_f=0$ and $N=4$.  
Also Eqs.(\ref{vzero},\ref{gzero}) represent a generalization of the flow 
equations for the simple Yukawa coupled fermion models derived in 
Refs.\cite{mythes,magg,clarka} to a model with both scalar and pseudoscalar 
bosons coupled to fermions.

\section{Numerical Results}
\label{num}

We now discuss the numerical solution of the LO $\sigma$-model RG flow 
equations, (\ref{flowa}) and (\ref{chflows}).  As mentioned in Appendix 
B, these equations is quite lengthy.
  Since we keep only ${\cal O}(\sigma)$ (first order in the chiral 
symmetry breaking parameter) in Eq. (\ref{sigpots}), there are six 
coupled flow equations, which for 
\begin{eqnarray}
     {\rm det}\ {\bf \Sigma}^{'}&=&F_0(\rho)+\sigma F_1(\rho)
          \nonumber \\
     {\rm tr}\ {\bf \Sigma}^{'-1}\cdot{\bf \Omega}&=&\Delta m_0(\rho)
          +{\cal G}\Delta g_0(\rho)+\sigma\biggl[\Delta m_1(\rho)
               +{\cal G}\Delta g_1(\rho)\biggr]
\end{eqnarray}
take the form:
\begin{eqnarray}
     \Lambda{\partial V_0\over{\partial\Lambda}}&=&-{A_d\over{2}}\Lambda^d
     \biggl({\rm ln}\ F_0(\rho)-n_fc_d\ {\rm ln}\ D_F\biggr)  
     \nonumber \\ 
     \Lambda{\partial V_1\over{\partial\Lambda}}&=&-{A_d\over{2}}\Lambda^d
     {F_1(\rho)\over{F_0(\rho)}}\label{flows} \nonumber \\  
     \Lambda{\partial m_0\over{\partial\Lambda}}&=&-{A_d\over{2}}\Lambda^d
          \Delta m_0(\rho)  
     \label{sigflows}\\ 
     \Lambda{\partial m_1\over{\partial\Lambda}}&=&-{A_d\over{2}}\Lambda^d
          \Delta m_1(\rho)\nonumber \\  
     \Lambda{\partial g_0\over{\partial\Lambda}}&=&-{A_d\over{2}}\Lambda^d
          \Delta g_0(\rho)\nonumber \\ 
     \Lambda{\partial g_1\over{\partial\Lambda}}&=&-{A_d\over{2}}\Lambda^d
          \Delta g_1(\rho). \nonumber 
\end{eqnarray}
Symbolic computing (we used {\it Mathematica}) greatly facilitates 
the determination of the functions 
$F_0$, $F_1$, $\Delta m_0$ etc. which are too lengthy to be reproduced 
here $\cite{getdetails}$. ($\Delta g_1(\rho)$ and $\Delta m_1(\rho)$, 
for instance, require several 
pages of output!).  Once these functions are determined, however, the 
numerical solution of Eqs.(\ref{sigflows}) proceeds exactly as in the scalar 
only case as reported, {\it e.g.}, in Refs \cite{hh,shepa} except with six 
functions instead of two.  There are a number of nonphysical subtleties associated 
with  the numerical solution  
of Eq.(\ref{sigflows}) (e.g. the domain limits, the algorithm used to solve
the equations, and the 
number of terms retained in the polynomial fits , etc.) 
These nonphysical sensitivities yield a small spread of 
output values for a given input. At present we have three
independently constructed codes to solve the equations which give 
essentially the same results which gives us some confidence in the 
solutions presented here.

We set our ``initial condition'' by specifying the value of the parameters 
at the UV scale:
\begin{eqnarray}
     V^{(\Lambda_0)}(\rho,\sigma)&=&{1\over2}\mu^2_0\rho^2+
     {1\over4}\lambda_0\rho^4 \nonumber \\
     U^{(\Lambda_0)}(\rho,\sigma,{\cal G})&=&m_q^0+g_0{\cal G}.
\end{eqnarray}
Then we expand in powers of $\rho$,
\begin{eqnarray}
     V_k^{(\Lambda)}(\rho)&=&\sum_{i=1}^M{1\over{2i}}
          v^{(2i)}_k(\Lambda)\rho^{2i} 
          \nonumber \\
     m_k^{(\Lambda)}(\rho)&=&\xi^{(0)}_k(\Lambda)+\sum_{i=1}^M{1\over{2i}}
          \xi^{(2i)}_k(\Lambda)\rho^{2i} 
          \nonumber \\
     g_k^{(\Lambda)}(\rho)&=&y^{(0)}_k(\Lambda)+\sum_{i=1}^M{1\over{2i}}
          y^{(2i)}_k(\Lambda)\rho^{2i}, 
\end{eqnarray}
for $k=0,1,2$ (see Eq.(\ref{sigpots})).  Thus at the UV scale we have
\begin{eqnarray}
     v^{(2)}_0(\Lambda_0)&=&\mu^2_0 \nonumber \\
     v^{(4)}_0(\Lambda_0)&=&\lambda_0 \nonumber  \\
     \xi^{(0)}_0(\Lambda_0)&=&m^0_q \\
     y^{(0)}_0(\Lambda_0)&=&g_0,
\end{eqnarray}
with all higher order coefficients for $k=1,2$ set to zero.  Just as in the 
previous section we perform a fit of the functions $V_k$, $m_k$, and $g_k$ 
to a power series in $\rho$ at each $\Lambda$ step.  Also, since we're 
interested in spontaneous symmetry breaking, we will have the parameter 
$f_\pi=<\phi >_{vac}=\sigma$, which sets the scale of the symmetry breaking.  
Thus we consider the set \{$m_\pi$, $f_\pi$, $\mu_0^2$, $\lambda_0$, $m^0_q$,  
$g_0$, $\Lambda_0$, $\Lambda_{IR}$\} the input parameters to the model. This 
approach is philosophically different from that taken by Jungnickel and 
Wetterich \cite{jungwett} where $f_\pi$ was the quantity they wished to predict. 

The basic approach adopted in the present work for constraining the phenomenology 
is as follows.  Since 
there are as yet no experimental or reliable theoretical constraints at the UV 
scale, we are forced to look to observables at the IR scale to determine the 
free parameters of the model.  
Perhaps the two parameters in the model most tightly constrained by low energy 
data are $m_\pi$ and $f_\pi$.  Thus we will tune other parameters of the model 
to get $m_\pi$ and $f_\pi$ at their experimental values at the infrared scale.  
This still leaves four out of the six parameters unconstrained, and
 we must decide which out of \{$\mu_0^2$, $\lambda_0$, $m^0_q$,  $g_0$, 
$\Lambda_0$, $\Lambda_{IR}$\} to fix and which to tune to $m_\pi$ and $f_\pi$.  
Results for scalar-only calculations \cite{hh,shepa} as well as for Yukawa 
coupled fermions \cite{mythes,magg,clarka} indicate that the two parameters 
$\lambda_0$ and $\mu_0^2$ are not truly independent, {\it i.e.} we can tune to 
particular values of IR parameters with large number of values for $\lambda_0$ 
and $\mu_0^2$ so long as $\lambda_0>0$ and large.  Thus we fix 
$\lambda_0=10$ and use 
$\mu_0$ to tune the IR parameters. From our experience choosing a different value of
$\lambda_0$ will only require a re-tuning of $\mu_0$ to obtain equivalent results. 
Next there is the bare current quark mass, $m_q^0$.  
Experimentally, there is about $\pm 5$MeV spread of the values for $m_u$ and 
$m_d$ \cite{pdg}. 
 We will tune $m_q^0$ to get the average of the means of these values at the 
IR scale, {\it i.e.} $m_0^{(\Lambda_{IR})}\equiv m_q={1\over{2}}(\overline 
m_u+\overline m_d)=7.5$MeV.  We also have the two cut-off scales $\Lambda_0$ and 
$\Lambda_{IR}$, only one of which need be tuned with the other fixed since the 
RG equations only care about the 
ratio.  We fix $\Lambda_{IR}=m_{\pi^{\pm}}\simeq 140$MeV and will allow 
$\Lambda_0$ to be adjusted.    This leaves 
only $g_0$.  We know that the quark-meson coupling is strong at these energy 
scales and the Goldberger-Treiman relation for the constituent quark model 
gives $g\simeq M_{nucleon}/3f_\pi\simeq 3.366$.  But this is not an approrpiate 
constraint in our zero-density model since we have no nucleons.  
We therefore perform three 
fits, one each for $g>,\simeq,<3.366$ and compare the results.  To summarize, for 
each of these values of $g_0$ we tune \{$\mu_0$, $m_q^0$, $\Lambda_0$\} to 
\{$m_\pi$, $f_\pi$, $m_q$\}.  Once all the parameters are fixed, the 
model predicts other IR quantities such as $m_\sigma$ (sigma mass), 
$\lambda_{4-pt}$, $\lambda_{3-pt}$ (the sigma four- and three-point couplings), 
$g$, $a_0^0$, and $a_0^2$.  Not all these quantities have experimental 
determinations however; the quantities best determined by 
experiment are the $\pi\pi$ scattering lengths $a_0^0$ and $a_0^2$ discussed 
in more detail later (see Appendix C for background).

The results  of the three fits for $g>,\simeq,<3.366$ are displayed in 
Table \ref{gfits}.  The middle column is the result for fitting $g_0$ to give
$g\simeq3.36$; the right and left columns are the results for arbitrarily 
choosing $g_0=2.500$ and $3.100$ respectively.  The last three rows 
(in the top section) contain the actual fits to $m_\pi$, $f_\pi$, and $m_q$.  
The point was to get ($m_\pi$, $f_\pi$, $m_q$) $\simeq$ ($140$MeV, $92.4$MeV,
 $7.5$MeV) \cite{pdg}.  The $\P$ and $\|$ indicate that $m_q^0$ and ($\mu_0$,
$\Lambda_0$) were used to fix $m_q$ and ($m_\pi$,$f_\pi$).  The values of 
$m_q^0$ and $m_q$ all fall within the uncertainty quoted in the particle data 
book \cite{pdg}.  Rows $9$ through $15$ represent some of the predictions of 
the calculation.  Since the sigma is less a ``particle'' and more a broad 
resonance, the values for $m_\sigma$, $\lambda_{4-pt}$, and $\lambda_{3-pt}$ 
are hard to compare quantitatively with experiment.  The values for the 
scalar density, $<\overline\psi\psi>$ are close to other calculated 
values of $<\overline\psi\psi>\sim-[(240\pm25){\rm MeV}]^3$
 \cite{njlcond}.  In addition
the adjusted values of $\Lambda_0\sim 940$MeV are in the range expected.

\begin{figure}
\centerline{\psfig{figure=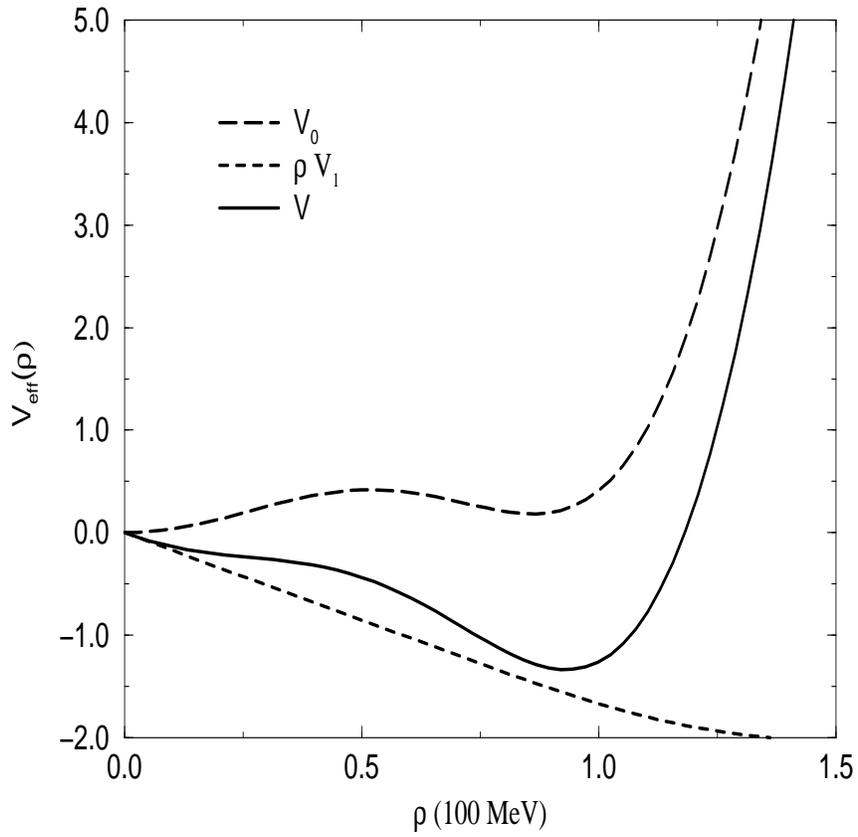,height=5in,width=5in}}
\caption{RG boson potentials for the $\sigma$-model.  The parameters 
of the calculation are displayed in Table 3.1}
\label{v0v1}
\end{figure}

\begin{figure}
\centerline{\psfig{figure=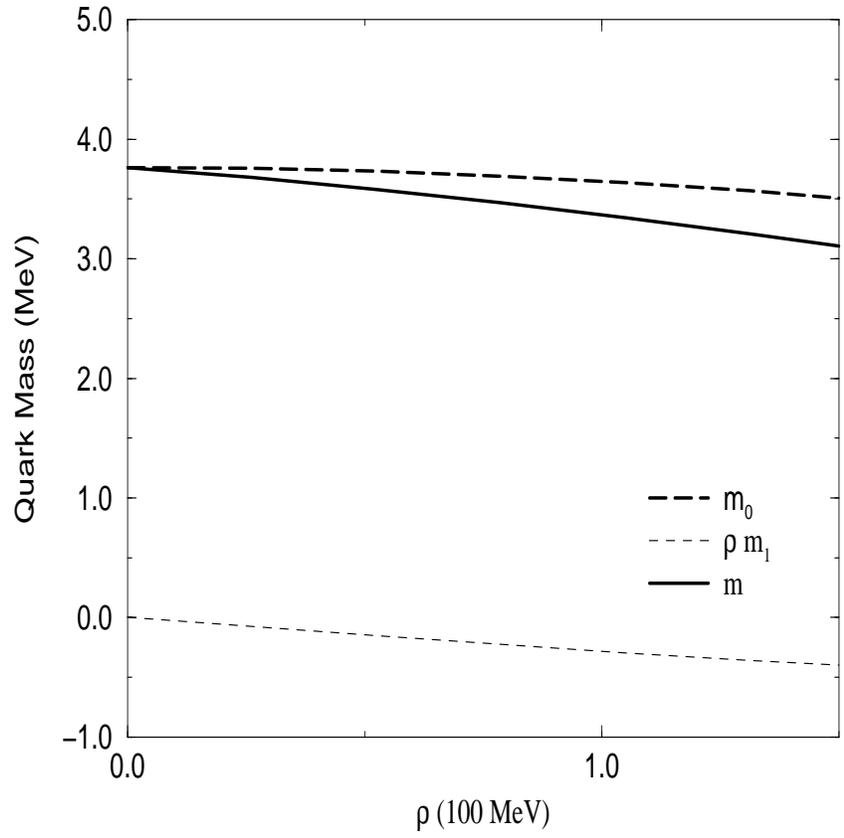,height=5in,width=5in}}
\caption{Quark mass funcions for the $\sigma$-model.  The parameters 
of the calculation are displayed in Table 3.1}
\label{mass}
\end{figure}

\begin{figure}
\centerline{\psfig{figure=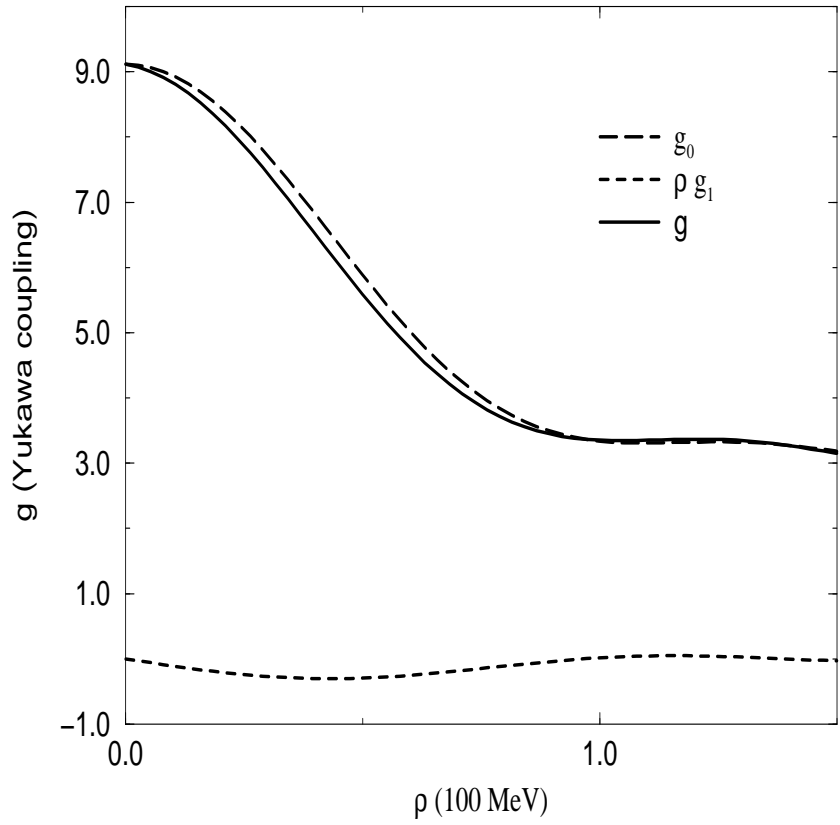,height=5in,width=5in}}
\caption{Yukawa coupling funcions for the $\sigma$-model.  The parameters 
of the calculation are displayed in Table 3.1}
\label{gyuk}
\end{figure}

Fig. \ref{v0v1} shows the boson potentials as a function of $\rho$.  As 
expected for weakly broken chiral symmetry, the first order term, $V_1$ 
is just a small correction.  One can see clearly that the minimum is at 
$\rho=<\sigma >_{vac}=f_\pi\simeq 93$ MeV \cite{pdg}.  Figs. \ref{mass} and 
\ref{gyuk} 
display the other functions computed in the model.  At $\rho\simeq 93$ 
MeV the values of $m$ and $g$ are just the IR values quoted in Table \ref{gfits}.  
The relatively small contribution from $g_1$ and $m$ justify {\it ex post facto} 
our expansion in the (small) chiral symmetry breaking parameter $\sigma$ (Eq.(\ref{sigpots})).

\section{$\pi\pi$ Scattering Lengths}

Another set of predictions from the RG solution of the $\sigma$-model 
come in the form of the parametrization of low energy $\pi\pi$ scattering.  
The expansion of the real part of the partial-wave amplitude can be written as 
\begin{equation}
     {\rm Re}\ A^I_l(s)=32\pi\ \biggl({q^2\over{m_\pi^2}}\biggr)^l
          \biggl(a^I_l+b^I_l{q^2\over{m_\pi^2}}+\cdots\biggr),
\end{equation}
where $I=0,1,2$ denotes the isospin channel and $l$ is the partial wave index. 
(See, {\it e.g.}, Refs.\cite{dynsm} section VI-4 and \cite{nelson}. 
 Appendix C contains a review of the perturbative 
calculation and the connection to our model.)
So for low energy ($q^2\ll m_\pi^2$) scattering $a^I_l$ and $b^I_l$ will 
be the most relevant quantities to study.  Table \ref{scattlen} displays a 
comparison of our three fits and a number of different calculations and 
experimental values for $a^0_0$ and $a^2_0$ in dimensions of inverse pion mass.  
Also, the quantity $2a^0_0-5a^2_0$ is included since for $s$-wave 
scattering it provides a constraint \cite{weinpi}.  Each of the three 
fits give results that are consistent with experiment and with $\chi$PT 
and lattice QCD calculations.  We discuss how the $\pi\pi$ scattering 
lengths are computed in our model in Appendix C.

\section{Summary and Possible Future Extensions of the Model}
\label{con}

We present a nonperturbative solution of the $\sigma$-model using the 
a sharp-cutoff RG equation truncated to LO in the derivative expansion.  The model
mimics the chiral symmetry of QCD. An important feature of our approach is that 
we can straight-forwardly track the chiral-symmetry-breaking. 
 Including a small quark mass at 
the UV scale,  we can follow the chiral symmetry breaking in the 
numerical solution as the scale is lowered.  The parameters  of 
the model can be constrained by low-energy data allowing predictions 
of, for example, $\pi-\pi$ scattering lengths.   The 
values for $\pi\pi$ scattering lengths obtained in this way 
show an improvement over the 
perturbative calculation \cite{weinpi} and are essentially consistent 
with experiment \cite{piexp} and other nonperturbative calculations 
\cite{pichpt,pilatt}.  It might be added that these calculations were 
performed at a substantially lower computational cost than those of 
Refs. \cite{pichpt,pilatt}.  This work contrasts with that of 
Ref. \cite{jungwett} in that we confine our attention to
the SU(2)xSU(2) sector, use a sharp cut-off approach truncated at LO, 
and allow for explict treatment of the chiral symmetry breaking 
via an expansion in the vacuum 
expectation value of the $\sigma$-field.  We also adopt a different philosophy as
regards the phenomenology constraining the free parameters.

There are a number of possible extensions to the present calculation. 
Perhaps the easiest is the extension to ${\cal O}(\sigma^2)$.  Indeed
much of the analytical work has already been done with the expected result that, 
for small current quark masses, the second order potentials will be small 
corrections to the first order, testifying to the convergence of the method.  
We have performed ${\cal O}(\sigma^2)$  calculations with only the bosonic 
potentials flowing ({\it i.e.} fixing $m(\rho)=m_q^0$ and $g(\rho)=g_0$ 
for all $\Lambda$) and confirmed 
that $V_2$ is small compared to $V_1$.  The results for the other 
calculated parameters in the model are not expected to change substantially at 
${\cal O}(\sigma^2)$ since $m_q^0$ is small.  

Another possible extension would be to allow for finite density which would permit
an analysis of nuclear phenomena.   Crudely, at finite density, there would be 
a momentum scale, $k_F$, below which the fermion parts of the flow 
equations would cease to contribute due to Pauli blocking, while the 
boson parts would still contribute to the flow.  Such crude calculations using 
this scheme show the qualitative restoration of chiral symmetry as $k_F$ 
increases.  A proper handling of finite density with the RG, however, 
for relativistic field theories requires much more care so we defer 
offering any conclusions at the present time.

Yet another extension is to incorporate strangeness into the model by 
extending to $SU(3)\times SU(3)$.  This requires the introduction of two 
more field invariants as
treated in Ref.\cite{jungwett}.  In this case, since 
$m_s\sim\Lambda_{QCD}$, chiral $SU(3)\times SU(3)$ is {\it strongly} 
broken and the expansion in the chiral symmetry breaking field may not converge 
rapidly enough and 
some other approach to following the chiral symmetry breaking may be required.

\begin{acknowledgements}

The authors gratefully acknowledge useful discussions with Veljko 
Dmitra\u sinovi\'c and Steve Pollock.  This work was supported in part 
by the U.S. DOE.
\end{acknowledgements}


\appendix

\section{Supermatrix Formalism}
\label{appendixa}

In this appendix we review the supermatrix formalism.  
A ``supermatrix'' is a square matrix of the form,
\begin{eqnarray}
     M=\pmatrix{M_{BB}&M_{BF} \cr 
     M_{FB}&M_{FF}},
\end{eqnarray}
where the square submatrices $M_{BB}$ and $M_{FF}$ are both {\it even} 
elements of the Grassmann algebra while $M_{BF}$ and $M_{FB}$ are {\it odd}
 elements of the Grassmann algebra.

(Grassmann or anticommuting 
variables allow the incorporation of Fermi statistics into the path integral 
formalism of quantum field theory.  An arbitrary function of a Grassmann variable
 can be Taylor expanded as 
\begin{equation}
     f(\theta)=\alpha+\beta\theta
\end{equation} where $\alpha$ and $\beta$ are normal numbers $\theta$ obeys 
the Grassmann algebra,
\begin{equation}\{\theta,\theta\}=\theta\theta+\theta\theta=0\Rightarrow\theta^2=0
\end{equation}
which is why the Taylor series terminates.  
For more details see , {\it e.g.}, 
Ref.\protect\cite{ramond} p.214-219.)

The supertrace is defined as
\begin{equation}
     {\rm str}M={\rm tr}M_{BB}-{\rm tr}M_{FF};
\end{equation}
so that the familiar commutative property of the normal trace still holds 
for the supertrace.  The superdeterminant is defined as
\begin{equation}
     {\rm sdet}M={\rm e}^{{\rm strln}M},
\end{equation}
which preserves the familiar property of determinants, ${\rm det}MN=
{\rm det}M{\rm det}N$.

Now consider the decomposition of M:
\begin{eqnarray}
     M=\pmatrix{M_{BB}&0\cr M_{FB}&1}
     \pmatrix{1&M^{-1}_{BB}M_{BF}\cr 0&N_{FF}}
\end{eqnarray}
where $N_{FF}=M_{FF}-M_{FB}M^{-1}_{BB}M_{BF}$.  Now it's easy to show
\begin{eqnarray}
     {\rm sdet}\pmatrix{M_{BB}&0\cr M_{FB}&1}={\rm det}M_{BB}
\end{eqnarray}
and,
\begin{eqnarray}
     {\rm sdet}\pmatrix{1&M^{-1}_{BB}M_{BF} \cr 0&N_{FF}}={\rm det}N_{FF}
\end{eqnarray}
so that,
\begin{equation}
     {\rm sdet}M={\rm det}M_{BB}{\rm det}^{-1}N_{FF}.
\end{equation}
Similarly we could have chosen the decomposition
\begin{eqnarray}
     M=\pmatrix{N_{BB}&M_{BF}M^{-1}_{FF} \cr 0&1}
     \pmatrix{1&0 \cr M_{FB}&M_{FF}},
     \label{appaa}
\end{eqnarray}
where $N_{BB}=M_{BB}-M_{BF}M^{-1}_{FF}M_{FB}$.  Then we would be led to
\begin{equation}
     {\rm sdet}M={\rm det}N_{BB}{\rm det}^{-1}M_{FF}.
     \label{appa}
\end{equation}

\section{Derivatives of the Chiral Functions}
\label{appendixb}

In the derivation of the flow equations for the $\sigma$-model, derivatives 
with respect to $\phi^a$ of functions of $\rho$, $\sigma$, and 
${\cal G}=\Gamma^a\phi^a$ are taken.  In this appendix we work out what 
these derivatives are.  Consider the potential
\begin{equation}
     U(\rho,\sigma,{\cal G})=m(\rho,\sigma)+{\cal G} g(\rho,\sigma)
\end{equation}
where,
\begin{equation}
     m(\rho,\sigma)=m_0(\rho)+\sigma m_1(\rho)+{\sigma^2\over{2}}m_2(\rho)
          +\cdots
\end{equation}
and similarly for g.  Now
\begin{eqnarray}
     {\partial U\over{\partial\phi^a}}&\equiv& U^{'a}
          =m^{'a}+{\cal G} g^{'a}+\Gamma^a g \nonumber \\
     {\partial^2 U\over{\partial\phi^a\partial\phi^b}}&\equiv& U^{''ab}
          =m^{''ab}+{\cal G} g^{''ab}+\Gamma^a g^{'b}+g^{'a}\Gamma^b.
\end{eqnarray}
We need to work out the derivatives of functions of $\rho$ and $\sigma$.  
Consider $f(\rho,\sigma)$, $f=m$ or $g$:
\begin{eqnarray}
     {\partial f\over{\partial\phi^a}}\equiv f^{'a}
          &=&f^{(0)'a}+\sigma f^{(1)'a}+{\sigma^2\over{2}}f^{(2)'a}
               +\delta^{a0}(f^{(1)}+\sigma f^{(2)})
     \nonumber \\
          &=&{\phi^a\over{\rho}}f^{'}+\delta^{a0}(f^{(1)}+\sigma f^{(2)})
     \nonumber \\
     {\partial^2 f\over{\partial\phi^a\partial\phi^b}}\equiv f^{''ab}
          &=&f^{(0)''ab}+\sigma f^{(1)''ab}+{\sigma^2\over{2}}f^{(2)''ab}
               +\delta^{a0}(f^{(1)'b}+\sigma f^{(2)'b})
     \nonumber \\
               &&+(f^{(1)'a}+\sigma f^{(2)'a})\delta^{0b}
               +\delta^{a0}\delta^{0b}f^{(2)}
     \nonumber \\
          &=&\delta^{ab}{f^{'}\over{\rho}}+{\phi^a\phi^b\over{\rho^2}}
               \biggl(f^{''}-{f^{'}\over{\rho}}\biggr)
     \nonumber \\
               &&+{\delta^{a0}\phi^b+\phi^a\delta^{0b}\over{\rho}}
               \bigg(f^{(1)'}+\sigma f^{(2)'}\biggl)
               +\delta^{a0}\delta^{0b}f^{(2)}.
     \label{fpp}
\end{eqnarray}
In the last of each of the above equalities we've used
\begin{eqnarray}
     h^{'a}&=&{\phi^a\over{\rho}}h^{'}(\rho) \nonumber \\
     h^{''ab}&=&\delta^{ab}{h^{'}\over{\rho}}+{\phi^a\phi^b\over{\rho^2}}
          \biggl(h^{''}-{h^{'}\over{\rho}}\biggr),
\end{eqnarray}
where $h=h(\rho)$, $h=m^{(i)}$ or $g^{(i)}$ for $i=0,1,2$.  Also
\begin{eqnarray}
     f^{'}(\rho)&=&f^{(0)'}+\sigma f^{(1)'}+{\sigma^2\over{2}}f^{(2)'}
     \nonumber \\
     f^{''}(\rho)&=&f^{(0)''}+\sigma f^{(1)''}+{\sigma^2\over{2}}f^{(2)''}.
\end{eqnarray}

With these derivatives, we can compute the matrices in section \ref{dersig}:
\begin{eqnarray}
     \Sigma^{ab}&\equiv&\Lambda^2\delta^{ab}+V^{''ab}(\rho,\sigma)
     \nonumber \\
     &=&\biggl(\Lambda^2+{V^{'}\over{\rho}}\biggr)\delta^{ab}
          +\biggl(V^{''}-{V^{'}\over{\rho}}\biggr)
               {\phi^a\phi^b\over{\rho^2}}
     \nonumber \\
     &&+\tilde V^{'}\ {\delta^{a0}\phi^b+\phi^a\delta^{0b}\over{\rho}}
          +V^{(2)}\delta^{a0}\delta^{0b}
     \nonumber \\
     &=&\Sigma_{\delta}\delta^{ab}
          +\Sigma_{\phi\phi}{\phi^a\phi^b\over{\rho^2}}
     +\Sigma_{\delta\phi}{\delta^{a0}\phi^b+\phi^a\delta^{0b}\over{\rho}}
          +V^{(2)}\delta^{a0}\delta^{0b}
     \label{bigsigab}
\end{eqnarray}
and
\begin{eqnarray}
     \Omega^{ab}&\equiv&U^{''ab}(\rho,\sigma,{\cal G})-{2\over{D_F}}
          U^{'a}(\rho,\sigma,{\cal G})U^{\dagger}(\rho,\sigma,{\cal G})
               U^{'b}(\rho,\sigma,{\cal G})
     \nonumber \\
     &=&{U^{'}\over{\rho}}\delta^{ab}+\biggl(U^{''}-{U^{'}\over{\rho}}
          -{2U^{\dagger}\over{D_F}}U^{'2}\biggl)
          {\phi^a\phi^b\over{\rho^2}}
     \nonumber \\
     &&+\biggl(\tilde U^{'}-{2U^{\dagger}\over{D_F}}U^{'}\tilde U\biggl)
          {\delta^{a0}\phi^b+\phi^a\delta^{0b}\over{\rho}}
     \nonumber \\
     &&+{2U^{\dagger}\over{D_F}}{\cal G} g^{'} \tilde m
               {\phi^a\delta^{0b}\over{\rho}}
     +{2U^{\dagger}\over{D_F}}{\cal G} m^{'} \tilde g
               {\delta^{a0}\phi^b\over{\rho}}
     \nonumber \\
     &&+\biggl(U^{(2)}-{2U^{\dagger}\over{D_F}}\tilde U^2\biggr)
          \delta^{a0}\delta^{0b}
     +\biggl(g^{'}-{2U^{\dagger}\over{D_F}}gU^{'}\biggr)
     {\Gamma^a\phi^b+\phi^a\Gamma^b\over{\rho}}
     \nonumber \\
     &&+\biggl(\tilde g-{2U^{\dagger}\over{D_F}}\tilde U\biggr)
          \biggl(\Gamma^a\delta^{0b}+\delta^{a0}\Gamma^b\biggr)
     -{2g^2\over{D_F}}\Gamma^aU^{\dagger}\Gamma^b
     \nonumber \\
     &=&\Omega_{\delta}\delta^{ab}+\Omega_{\phi\phi}
          {\phi^a\phi^b\over{\rho^2}}
     +\Omega_{\delta\phi}^S{\delta^{a0}\phi^b+\phi^a\delta^{0b}\over{\rho}}
     +\Omega_{\phi\delta}{\phi^a\delta^{0b}\over{\rho}}
     +\Omega_{\delta\phi}{\delta^{a0}\phi^b\over{\rho}}
     \nonumber \\
     &&+\Omega_{\delta\delta}\delta^{a0}\delta^{0b}
     +\Omega_{{\cal G}\phi}{\Gamma^a\phi^b+\phi^a\Gamma^b\over{\rho}}
     +\Omega_{{\cal G}\delta}(\Gamma^a\delta^{0b}+\delta^{a0}\Gamma^b)
     -\Gamma^a\Omega_{\cal G G}\Gamma^b,
     \label{omegab}
\end{eqnarray}
where
\begin{eqnarray}
     \tilde X&=&X^{(1)}+\sigma X^{(2)}, \qquad X=V,U,m,g
     \nonumber \\
     U^{'}&=&m^{'}+{\cal G} g^{'}.
\end{eqnarray}
Note that the $\Sigma$s are all functions of $\rho$ and $\sigma$ and the 
$\Omega$s are all functions of $\rho$, $\sigma$ and ${\cal G}$.  The 
ordering in all the terms containing ${\cal G}$s is nontrivial since 
the Pauli matrices don't commute with each other.

Now using the similarity transformation matrix (defining $\pi_1^{'2}=
\pi_1^2+\pi_2^2$ and $\pi_2^{'2}=\pi_2^2+\pi_3^2$),
\begin{eqnarray}
     {\bf S}={1\over{\rho}}\pmatrix{\sigma & \pi_1 & \pi_2 & \pi_3 \cr
               -\pi_1' & {\sigma\pi_1\over{\pi_1'}} 
               & {\sigma\pi_2\over{\pi_1'}} 
               & {\sigma\pi_3\over{\pi_1'}} \cr
               0 & -{\rho\pi_2'\over{\pi_1'}} &
               \rho{\pi_2\pi_1\over{\pi_2'\pi_1'}} &
               \rho{\pi_3\pi_1\over{\pi_2'\pi_1'}} \cr
               0 & 0 & -\rho{\pi_3\over{\pi_2'}} &
               \rho{\pi_2\over{\pi_2'}}},
\end{eqnarray}
(obtained by multiplying rotation matrices in $4$-space---see Appendix D 
in Ref. \cite{mythes}) we can compute,
\begin{eqnarray}
     {\bf \Sigma}^{'}&=&{\bf S}\cdot{\bf \Sigma}\cdot{\bf S}^T
     \nonumber \\
     {\bf \Omega}^{'}&=&{\bf S}\cdot{\bf \Omega}\cdot{\bf S}^T,
\end{eqnarray}
this amounts to performing similarity transformations on each of the tensors
\begin{eqnarray}
     \delta^{ab},\qquad {\phi^a\phi^b\over{\rho^2}}, \qquad
     {\phi^a\delta^{0b}\over{\rho}}, \qquad
     {\delta^{a0}\phi^b\over{\rho}}, \qquad
     \delta^{a0}\delta^{0b}, 
     \nonumber \\
     {\Gamma^a\phi^b+\phi^a\Gamma^b\over{\rho}}, \qquad
     (\Gamma^a\delta^{0b}+\delta^{a0}\Gamma^b), \qquad
     \Gamma^a\Gamma^b,
\end{eqnarray}
as can be read off from Eqs.(\ref{bigsigab}) and (\ref{omegab}).  After 
performing the similarity transformation ${\bf \Sigma}^{'}$ has the form
\begin{equation}
     {\bf \Sigma}^{'}=\pmatrix{\Sigma_{00} & \Sigma_{01} & 0 & 0 \cr
                    \Sigma_{10} & \Sigma_{11} & 0 & 0 \cr
                     0 & 0 & \Sigma_{22} & 0 \cr
                     0 & 0 & 0 & \Sigma_{33}},
\end{equation}
since $\Sigma^{01}=\Sigma^{10}$ we can write
\begin{eqnarray}
     {\rm det}\ {\bf \Sigma}^{'}&=&(\Sigma^{00}\Sigma^{11}-(\Sigma^{01})^2)
          \Sigma^{22}\Sigma^{33}
     \nonumber \\
     {\rm tr}\ {\bf \Sigma}^{'-1}\cdot{\bf \Omega}^{'}&=&
          (\Sigma^{00})^{-1}\Omega^{00}+
          (\Sigma^{10})^{-1}\Omega^{01}+
          (\Sigma^{01})^{-1}\Omega^{00}
     \nonumber \\
          &&+(\Sigma^{11})^{-1}\Omega^{11}+
          (\Sigma^{22})^{-1}\Omega^{22}+
          (\Sigma^{33})^{-1}\Omega^{33}.
     \label{traceomsigapp}
\end{eqnarray}
Thus only $12$ elements---$6$ from ${\bf \Sigma}^{'}$ and $6$ from 
${\bf \Omega}^{'}$---need be computed to determine the flow equations 
for the $\sigma$-model.  As is probably clear this substantially 
reduces the complexity of the expressions but they they are still 
quite complicated.  The derivation is facilitated by using Mathematica 
to compute the $\Sigma$ and $\Omega$ functions in Eqs.(\ref{bigsigab}) 
and (\ref{omegab}).  Then Eqs.(\ref{traceomsigapp}) can be written 
in terms of these expressions.

\section{$\pi\pi$ Scattering Lengths in the RG Model}
\label{appendixc}  

In this appendix we will sketch the calculation of $\pi\pi$ scattering lengths. 
 Before addressing the calculation in our model we first discuss the 
calculation using the perturbative $\sigma$-model.  This was first 
done by Weinberg \cite{weinpi}; useful reviews appear here as 
Refs. \cite{dinko,dynsm}.

Differential cross sections in field theory are computed by squaring 
the ``invariant amplitude'', $A$ which is usually computed to a given 
order in perturbation theory using diagrammatic techniques (for details 
on our conventions see {\it e.g.} Ref.\cite{itzub} Appendix A-3):
\begin{equation}
     {d\sigma\over{d\Omega}}\propto|A|^2.
\end{equation}
Consider the amplitude for processes involving two pions in and two 
pions out.

As indicated, it will depend on the isospin channel $I=0,1,2$, the 
Mandelstam momentum variables $s=(p_1+p_2)^2$, $t=(p_1+p_3)^2$, 
$u=(p_1+p_4)^2$ and the isospin indices of each of the pions, $i,j,k,l$. 
 (In the center-of-mass frame these become $s=4(m_\pi^2+\vec k^2)$,
$t=-2(1-\cos\theta)\vec k^2$, and $u=-2(1+\cos\theta)\vec k^2$, where 
$\vec k$ is the $3$-momentum of the incident pion and $\theta$ is the 
angle between the incident pion and the out going pion.)  We can expand 
the amplitude in partial waves
\begin{equation}
     A^I(s,t,u)=\sum_{l=0}^\infty(2l+1)A^I_l(s,t,u)P_l(\cos\theta),
\end{equation}
where $P_l(\cos\theta)$ are the Legendre polynomials.  Amplitudes for each 
of the isospin channels are not independent, however; since all the 
particles are bosons $A^I_{ijkl}(s,t,u)$ is totally symmetric in all 
indices.  This can be exploited to show that there is really only one
independent amplitude $A_l(s,t,u)$.  Each of the amplitudes $A^I_l$ 
can be written in terms of $A_l(s,t,u)$:
\begin{eqnarray}
     A^0_l(s,t,u)&=&3A_l(s,t,u)+A_l(t,s,u)+A_l(u,t,s) \\ \nonumber 
     A^1_l(s,t,u)&=&A_l(t,s,u)-A_l(u,t,s)  
     \label{crossing}\\
     A^2_l(s,t,u)&=&A_l(t,s,u)+A_l(u,t,s)\nonumber.
\end{eqnarray}
So we need only compute $A_l(s,t,u)$.

At tree level, or lowest order in the coupling constant $\lambda$, 
the calculation is quite simple.  The Feynman rules can be read off 
of the boson potential,
\begin{equation}
     V(\rho)={1\over{2}}\mu^2\rho^2+{1\over{4}}\lambda\rho^4=
          \cdots -\lambda \sigma_{min}\sigma\vec\pi^2-
          {\lambda\over{4}}\vec\pi^4+\cdots.
\end{equation}
We consider the process $\pi^+\pi^-\longrightarrow\pi^0\pi^0$ with $\pi_0=\pi^3$, $\pi_\pm={1\over{\sqrt{2}}}(\pi^1\pm\pi^2)$, then we have
\begin{equation}
     V(\rho)=\cdots -\lambda\sigma_{min}\sigma(\pi_0^2+2\pi_+\pi_-)
          -{\lambda\over{4}}(\pi_0^4+4\pi_+^2\pi_-^2+4\pi_0^2\pi_+\pi_-)
          +\cdots.
\end{equation}
The amplitude can be written as

\begin{eqnarray}
     A_l(s,t,u)&=&-2i\lambda+4(-i\lambda \sigma_{min}){i\over{s-m^2_\sigma}}
          \\ \nonumber
          &=&-2i\lambda\biggl(1+{{m_\sigma^2-m_\pi^2}\over{s-m_\sigma^2}}
               \biggl)
     \label{amp}
          \\ 
          &=&\biggl({{s-m_\pi^2}\over{f_\pi^2}}\biggr)
               \biggl({{m_\sigma^2-m_\pi^2}\over{m_\sigma^2-s}}\biggr)
          \simeq {{s-m_\pi^2}\over{f_\pi^2}}\nonumber
\end{eqnarray}
where we used $m_\pi^2\ll m_\sigma^2$ and $s=4m_\pi^2$ (at threshold) 
in the last approximation.  Also we used the relations
\begin{eqnarray}
     \sigma_{min}&=&-f_\pi \\ 
     \lambda&=&{{m_\sigma^2-m_\pi^2}\over{2f_\pi^2}}\\
     \mu^2&=&{1\over{2}}(m_\sigma^2-3m_\pi^2) 
\end{eqnarray}
to replace ($\sigma_{min},\lambda,\mu^2$) with the observable 
($m_\sigma,m_\pi,f_\pi$).  From Eq. (\ref{crossing}) we can compute 
the amplitudes for the isospin channels,
\begin{eqnarray}
     A^0_0(s,t,u)&=&3A(s,t,u)+A(t,s,u)+A(u,t,s) \\ \nonumber 
     A^1_1(s,t,u)&=&A(t,s,u)-A(u,t,s)  
     \label{crossinga}\\
     A^2_0(s,t,u)&=&A(t,s,u)+A(u,t,s)\nonumber.
\end{eqnarray}
using Eq.(\ref{amp}), $s+t+u=4m_\pi^2$, $s=4(m_\pi^2+\vec k^2)$
\begin{eqnarray}
     A_0^0&=&{{2s-m_\pi^2}\over{f_\pi^2}}=7{m_\pi^2\over{f_\pi^2}}
          +8{m_\pi^2\over{f_\pi^2}}{\vec k^2\over{m_\pi^2}} \\ \nonumber
     A_1^1&=&{{t-u}\over{f_\pi^2}}={4\over{3\pi}}
          {m_\pi^2\over{f_\pi^2}}{\vec k^2\over{m_\pi^2}} \\ 
     A^2_0&=&{{t+u-2m_\pi^2}\over{f_\pi^2}}=-2{m_\pi^2\over{f_\pi^2}}
          -{4\over{\pi}}{m_\pi^2\over{f_\pi^2}}{\vec k^2\over{m_\pi^2}}.
     \nonumber
\end{eqnarray}
The ``scattering length'' $a_l^I$ and ``slope parameter'' $b_l^I$ are defined by
\begin{equation}
     {\rm Re}\ A^I_l(s)=32\pi\ \biggl({q^2\over{m_\pi^2}}\biggr)^l
          \biggl[a^I_l+b^I_l{q^2\over{m_\pi^2}}+
          {\cal O}({\vec k^4\over{m_\pi^4}})+\cdots\biggr],
\end{equation}
which gives the tree level values \cite{weinpi,dynsm}
\begin{eqnarray}
     a_0^0={7\over{32\pi}}{m_\pi^2\over{f\pi^2}}\simeq 0.16&;&\qquad
     b_0^0={8\over{32\pi}}{m_\pi^2\over{f\pi^2}}\simeq 0.18 \\ \nonumber
     a_1^1={1\over{24\pi}}{m_\pi^2\over{f\pi^2}}\simeq 0.0 &;&\qquad
     b_1^1=0 \\ 
     a_0^2=-{1\over{16\pi}}{m_\pi^2\over{f\pi^2}}\simeq -0.045&;&\qquad
     b_0^2=-{1\over{16\pi}}{m_\pi^2\over{f\pi^2}}\simeq -0.09
     \nonumber
\end{eqnarray}

The calculation in our model proceeds similarly.  Since we compute 
the potential $V(\rho)$ we must relate the tree diagrams in the invariant 
amplitude to this potential.  In the leading order (LO) approximation, 
all the momenta on the external legs of the diagrams are zero and 
therefore we can only compute the $s$-wave ($l=0$) scattering lengths.

We equate the vertex between two pions and one sigma with the third 
derivative of the potential 
and the $4$-pion 
vertex with the fourth derivative.

  Computation of these is straightforward using Eq.(\ref{fpp}) as a 
starting point,
\begin{eqnarray}
     {\partial^3V\over{\partial\sigma\partial\pi^i\partial\pi^j}}
     \biggl|_{\stackrel{\vec\pi=0}{\rho,\sigma=f_\pi}}
          &=&\delta^{ij}\biggl({V^{''}\over{f_\pi}}-{V^{'}\over{f_\pi^2}}
          \biggr) \\ \nonumber
{\partial^4V\over{\partial\pi^i\partial\pi^j
          \partial\pi^k\partial\pi^l}}
     \biggl|_{\stackrel{\vec\pi=0}{\rho,\sigma=f_\pi}} 
          &=&(\delta^{ij}\delta^{kl}+\delta^{ik}\delta^{jl}
          +\delta^{il}\delta^{jk})
          \biggl({V^{''}\over{f_\pi^2}}-{V^{'}\over{f_\pi^3}}
          \biggr)
\end{eqnarray}
where
\begin{equation}
     V^{'}=V^{'}_0(\rho)+\sigma V_1^{'}(\rho)+{\sigma\over{2}}V^{'}_2(\rho)
\end{equation}
and similarly for $V^{''}$.  Defining 
$F(\rho)={V^{''}\over{f_\pi}}-{V^{'}\over{f_\pi^2}}$ and 
$G(p)=F(\rho)/f_\pi+F^2(\rho)/(p-m_\sigma^2)$ we can construct 
the invariant amplitude,
\begin{eqnarray}
     A(s,t,u)&=&\delta^{ij}\delta^{kl}G(s)+\delta^{ik}\delta^{jl}G(t)
          +\delta^{il}\delta^{jk}G(u).
\end{eqnarray}

At threshold $s=4m_\pi^2$, $t=u=0$, so
\begin{equation}
     A(s)=\delta^{ij}\delta^{kl}\biggl({F(\rho)\over{f_\pi}}
          +{F^2(\rho)\over{4m_\pi^2-m_\sigma^2}}\biggr).
\end{equation}
Thus $A(t,s,u)=A(u,t,s)=A(0)$ and our crossing relations (Eq.(\ref{crossinga})) give,
\begin{eqnarray}
     A_0^0&=&3A(s)+2A(0) \\ \nonumber
     A^2_0&=&2A(s)
\end{eqnarray}
from which we have,
\begin{eqnarray}
     a_0^0&=&{1\over{32\pi}}A_0^0={1\over{32\pi}}(3A(s)+2A(0)) \\ \nonumber
     a^2_0&=&{1\over{32\pi}}A_2^0={1\over{16\pi}}A(s).
\end{eqnarray}


\vfill\eject


\begin{table}
\caption{Results for numerical solution of $\sigma$-model RG flow equations 
with 6 quark flavors for three different values of the UV quark-meson coupling $g_0$.
The $\P$ and $\|$ indicate that $m_q^0$ and ($\mu_0$,
$\Lambda_0$) were used to fix $m_q$ and ($m_\pi$,$f_\pi$) respectively.}
\begin{tabular}{cccc}
Parameter & Fit 1 & Fit 2 & Fit 3 \\ 
\hline
$g$                    & 2.967     & 3.358     & 3.893     \\ 
$g_0$                    & 2.500     & 2.765     & 3.100 \\ 
$\mu_0$(MeV)$^{\|}$          & 666.0     & 739.4     & 818.0     \\ 
$m_q^0$(MeV)$^{\P}$               & 6.54     & 6.42     & 6.36     \\ 
$\Lambda_0$(MeV)$^{\|}$          & 950.0     & 937.9     & 927.0     \\ 
$m_\pi$(MeV)$^{\|}$          & 140     & 140     & 140     \\ 
$f_\pi$(MeV)$^{\|}$          & 92.60     & 92.49     & 92.44     \\ 
$m_q$(MeV)$^{\P}$               & 7.48     & 7.51     & 7.54     \\ 
\hline \hline
$m_\sigma$(MeV)               & 507.3     & 536.1     & 550.4     \\ 
$\lambda_{4-pt}$          & 17.3     & 20.8     & 30.1     \\ 
$\lambda_{3-pt}$(100 MeV)        & 43.0     & 50.8     & 64.2     \\ 
$<-\overline\psi\psi>^{1/3}$(MeV)          &190     & 195     & 201     \\ 
$a_0^0(m_\pi^{-1})$          & 0.232     & 0.225     & 0.220     \\ 
$a_0^2(m_\pi^{-1})$          &-0.042     &-0.043     &-0.043     \\ 
$(2a_0^0-5a_0^2)(m_\pi^{-1})$     & 0.677     & 0.664     & 0.656     \\ 
\end{tabular}
\label{gfits}
\end{table}


\begin{table}
\caption{Comparison of $s$-wave $\pi\pi$ scattering lengths obtained by 
measurement and various calculations in dimensions of inverse pion mass.  
Experimental results in the first row are from Ref.\protect\cite{piexp}.  
The calculations in the second through the third row are quoted from 
Refs.\protect\cite{weinpi,pichpt,pilatt} respectively and were performed 
using the perturbative $\sigma$-model, chiral perturbation theory ($\chi$PT), 
and lattice QCD respectively.  In the last three rows are the results from 
our model for the three fits used in Table \protect\ref{gfits}.  Comprehensive 
reviews of $\pi\pi$ scattering are given in Ref.\protect\cite{dynsm} 
section VI-4 and Ref.\protect\cite{dinko}.}
\begin{tabular}{cccc}
&$a^0_0(m_\pi^{-1})$&$a^2_0(m_\pi^{-1})$&$2a^0_0-5a^2_0$ \\
\hline
Experiment    &$0.26\pm0.05$     & $-0.028\pm0.012$     & $0.66\pm0.12$     \\
Pert. 
$\sigma$-model     & $0.16$     & $-0.045$          & $0.56$     \\
$\chi$PT     & $0.20$     & $-0.042$          & $0.65$     \\
Lattice QCD      & $0.22$     & $-0.042$          & $0.65$     \\
Fit 1          & $0.232$     & $-0.42$          & $0.677$     \\
Fit 2          & $0.225$     & $-0.043$          & $0.664$     \\
Fit 3          & $0.220$     & $ -0.043$          & $0.656$     \\
\end{tabular}
\label{scattlen}
\end{table}

%
\vskip .5in


\vskip .5in


\vskip .5in


\begin{references}
\bibitem[\ddagger]{getdetails} See Ref.\cite{mythes} for full details.  
         Soft copies of the expressions are available from the authors.
\end{references}
\end{document}